\documentstyle[12pt]{article}
\include{epsf}
\include{psfig}
\input{psfig.sty}

\begin{document}

\parskip 18pt
\baselineskip 0.2in

\title{Measuring information spatial densities}

\author{Michele Bezzi$^(1)$, In\'es Samengo$^(1)$, \\ Stefan
Leutgeb$^(2)$ and Sheri J. Mizumori$^(3)$}

\date{$^(1)$Cognitive Neuroscience Sector, S.I.S.S.A, Via Beirut 2-4,
34014, Trieste, Italy \\
$^(2)$Program in Neuroscience and Department of Psychology,
University of Utah, Salt Lake City, Utah 84112\\
$^(3)$ Psychology Department, University of Washington, Seattle,
WA 98195}

\maketitle

\begin{center}
{\bf Abstract}

\parbox{10cm}{
A novel definition of the stimulus-specific information is
presented, which is particularly useful when the stimuli
constitute a continuous and metric set, as for example, position
in space. The approach allows one to build the spatial
information distribution of a given neural response. The method
is applied to the investigation of putative differences in the
coding of position in hippocampus and lateral septum.}
\end{center}

\vspace*{1cm}


\section{Introduction}

It has longly been known that many of the pyramidal cells in the
rodent hippocampus selectively fire when the animal is in a
particular location of its environment (O'Keefe and Dostrovsky,
1971). This phenomenon gave rise to the concept of place fields
and place cells, that is to say, the association between a given
cell and the particular region of space where it fires. Many
computational models of hippocampal coding for space (Tsodyks,
1999) are based on the idea that the information provided by each
place cell concerns whether the animal is or is not at a
particular location---the place field. It is clear, however, that
at least in principle, it could be possible that a given cell
provided an appreciable amount of information about the position
of the animal without having a place field in the rigorous sense.
For example, it could happen that a cell indiscriminately fired
all over the environment {\it except} in one specific location,
where it remained silent. Such a coding would be particularly
informative in those occasions where the neuron did not fire.
More generally, a cell could fire throughout the whole
environment but with a stable and reproducible distribution of
firing rates strongly selective to the position.  Place field
coding would mean that such a firing distribution is a very
specific one, namely, one where the cell remains silent
everywhere, except inside the place field. In this sense, the
idea of a place cell is to the coding of position what in other
contexts has been referred to as a grandmother-cell. Whether a
given place cell behaves strictly as a grandmother-cell or not,
depends on the size of the spatial bins. However, broadly
speaking, place cells use a sparse code and only respond to a
very small fraction of all possible locations.

\begin{figure}[htbf]
\begin{center}
\leavevmode\epsfxsize=10.00truecm \epsffile{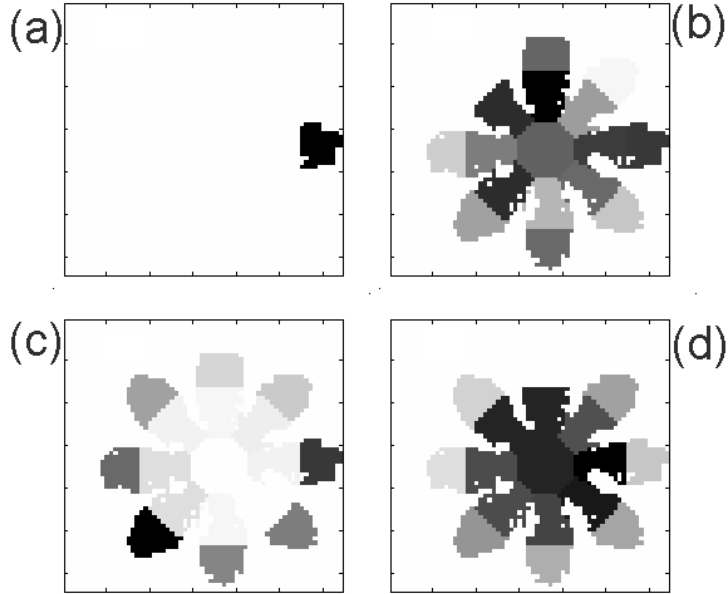}
\end{center}
\caption{Firing rate distribution of four different neurons when
a rat is exploring an 8-arm maze. In each case, the density of
the color is proportional to the number of spikes per unit time,
as a function of space. Since cells of different types have very
dissimilar mean firing rates, each plot has been normalized. The
absolute rates are shown in table \ref{t1}. Each cell provides an
amount of information which is at least as large as $\langle I
\rangle + \sigma_I$, where $\langle I \rangle$ is the mean
information provided by the whole set of cells of that particular
type, and $\sigma_I$ is its standard deviation (see Table
\ref{t1a} in Section 3 for the quantitative details).} \label{f0}
\end{figure}

In Figure \ref{f0}
we show four examples of the firing rate distribution of four
different neurons when a rat is exploring an 8 arm maze. Similar
to previous reports (Zhou {\it et al.} 1999), all of these
neurons provide an appreciable amount of information about the
location of the animal (see caption of Figure \ref{f0}).
\begin{table}[htbf]
\begin{center}
\begin{tabular}{||c|l|r|r||} \hline
case & type of neuron &  $\langle r \rangle$ & $I/t$ \\
& & spikes / sec. &  bits / sec. \\ \hline a & hipp. pyramidal
cells & 0.1981 & 0.7431 \\ \hline b & hipp. interneuron & 15.8995
& 2.1941 \\ \hline c & lat. septum &
2.2688 & 1.2921 \\ \hline d & med. septum & 10.4878 & 0.4411 \\
\hline
\end{tabular}
\end{center}
\caption{Data corresponding to the cells whose firing density is
shown in Figure \ref{f0}. $\langle r \rangle$ indicates the mean
firing rate of the cell, averaged throughout the maze, while
$I/t$ is the (corrected) information rate (see equations
(\ref{limsam}) and (\ref{stl})).} \label{t1}
\end{table}
Table \ref{t1} summarizes the data corresponding to the figure.
Chart (a) shows a typical place field, the cell only fires when
the animal reaches the end point of the right arm. In (b) and (c)
we show two  different distributed codes, the first corresponds to
an hippocampal interneuron and the latter to a neuron in the
lateral septum that selectively fires when the rat occupies the
endpoints of the maze. Finally, (d) shows a cell in the medial
septum whose discharge  corresponds to all locations within the
environment, with a somewhat lower rate in the endpoints of the
maze, specially in the upper arm. These examples show that there
might be different coding schemes for position, other than
localized place fields. Here, we explore such coding strategies,
both in hippocampal pyramidal neurons and in lateral septal cells
of behaving rats. The latter cell type receives a massive
projection from the hippocampus (Swanson {\it et al.} 1981, Jakab
{\it et al.} 1995), which presumably provides information about
spatial location. Our aim is to see whether different types of
neurons use different codes to represent position.

In the following section, the local information is defined, and its
relation to previous similar quantities is established. By taking the
limit of a very fine binning, such a local information gives rise to a
spatial information density, that can be used to explore the coding
strategy of a cell. In Section 3 the spatial information distribution
is calculated for actual recordings in rat hippocampal and lateral
septal cells. We end in Section 4, with some concluding remarks.

\section{Stimulus-specific informations}

Our analysis is based on the calculation of the mutual information between
the neural response of each single cell and the location of the animal
\begin{equation}
I =
\sum_{j} \sum_{n} P(n, {\bf x}_j) \
\log_2\left[ \frac{P(n, {\bf x}_j)}{P(n) P({\bf x}_j)} \right],
\label{inf}
\end{equation}
where ${\bf x}_j$ is a small region of space, and $n$ is the number
of spikes fired in a given time window. $P(r, {\bf x}_j)$ is the joint
probability of finding the rat at ${\bf x}_j$ while measuring response $n$,
and can always be written as $P(n, {\bf x}_j) = P(n | {\bf x}_j)
P({\bf x}_j)$. The a priori probability $P({\bf x}_j)$ is estimated
from the ratio between the time spent at position ${\bf x}_j$ and
the total time. The probability of response $n$ reads
\begin{equation}
P(n) = \sum_{j} P(n, {\bf x}_j).
\end{equation}
The mutual information $I$ measures the selectivity of the firing
of the cell to the location of the animal. It quantifies how much
can be learned about the position of the rat by looking at the
response of the neuron. In contrast to other correlation
measures, its numerical value does not depend on whether the cell
only fires in a particular location, or whether it only remains
silent there. It may happen, however, that the neural responses
are highly selective to some very specific locations, and not to
others. It is clear that the quantity defined in (\ref{inf})
provides the total amount of information, averaged over all
positions. The scope of the present section is to characterize
the detailed structure of the spatial locations where the cell is
most informative. To do so, we would like to build a spatial
information map; that is a way  to quantify the amount of
information provided by the cell about every single location
${\bf x}_j$. This issue has been discussed in De Weese and
Meister (1999), although in the context of more general stimuli,
not specifically position in space. Two definitions (among
infinitely many) have been pointed out, namely, ``the stimulus
specific surprise'' (which in the present case will be addressed
as the {\it position} specific surprise)
\begin{equation}
I_{1}({\bf x}_j) = \sum_n P(n | {\bf x}_j) \log_2 \left[ \frac{P(n |
{\bf x}_j)}
 {P(n)} \right],
\label{ses}
\end{equation}
and the ``stimulus specific information'' ({\it position}
specific
 information)
\begin{equation}
I_2({\bf x}_j) =  - \sum_n P(n) \log_2 \left[P(n)\right]
 +
 \sum_n P(n | {\bf x}_j) \log_2 \left[ P(n | {\bf x}_j)
\right].
\label{sei}
\end{equation}
Both of these quantities, when averaged in ${\bf x}_j$, give the total
information
 (\ref{inf})
\begin{equation}
\sum_{j} P({\bf x}_j) I_{1,2}({\bf x}_j) = I.
\label{parc}
\end{equation}
However, none of the two is, by itself, a proper information. The
stimulus specific surprise (\ref{ses}) is guaranteed to be
positive, but may not be additive, while the stimulus specific
information (\ref{sei}) is additive, but not always positive.
Moreover, any weighted sum of $I_1$ and $I_2$ is also a valid
estimator of the information to be associated to each location
(De Weese and Meister, 1999). However, in specific situations
these two local information estimators can be very different,
which means that their weighted sum can, in practice, lead to any
possible result.

Let us examine the behavior of $I$ and $I_{1,2}$ in the short
time limit. We consider a time interval $t$, and a cell whose
mean firing rate at position ${\bf x}_j$ is $r({\bf x}_j)$.
Therefore, if $t \ll 1 / r({\bf x}_j)$ the cell will most
probably remain silent at ${\bf x}_j$, only seldom firing a
spike. The short time approximation involves discarding any
response consisting of two or more spikes. Rigorously speaking,
it does not mean that such events will not occur, but rather,
that the set of symbols that are considered as informative
responses are the firing of a single spike, with probability $P(1
| {\bf x}_j) \approx r({\bf x}_j) t$, and whatever other
event---which we call response 0--- with probability $P(0 | {\bf
x}_j)= 1 - P(1 | {\bf x}_j)$. Therefore, as derived in Skaggs
{\it et al.} (1993) and Panzeri {\it et al.} (1996),
\begin{equation}
I =  t \sum_{j} P({\bf x}_j) \left\{ r({\bf x}_j) \log_2 \left[
\frac{r({\bf x}_j)}{\langle r \rangle}
\right] + \frac{\langle r \rangle - r({\bf x}_j)}{\ln 2} \right\} +
{\cal O}(t^2), \label{it}
\end{equation}
where
\begin{equation}
\langle r \rangle = \sum_{j} P({\bf x}_j) r({\bf x}_j).
\label{stl}
\end{equation}
The short time limit of $I$ is much more easily evaluated
from recorded data than the full equation (\ref{inf}), since
it does not need the estimation of the conditional probabilities.
Only the firing rates at each location are needed.
The first term in the curled brackets comes from the firing of a spike
in ${\bf x}_j$, while the second describes the silent response.

Similarly, $I_{1,2}$ tend to
\begin{eqnarray}
I_1({\bf x}_j) &=& t \left\{ r({\bf x}_j) \log_2\left[ \frac{r({\bf x}_j)}
{\langle r \rangle} \right]
 + \frac{\langle r \rangle - r({\bf x}_j)}{\ln 2} \right\} + {\cal O}(t^2),
\label{stli1}  \\
I_2({\bf x}_j) &=&  t \left\{\frac{\langle r \rangle - r({\bf x}_j)}
{\ln 2} + r({\bf x}_j) \log_2[r({\bf x}_j) t] - \langle r \rangle
\log_2(\langle r \rangle t)\right\} + {\cal O}(t^2). \label{stli2}
\end{eqnarray}
Equation (\ref{stli1}) states that the stimulus specific surprise also
rises linearly as a function of $t$. Its first term comes from those
cases when the cell fires a spike, while the second corresponds to the
silent response. The stimulus specific information, on
the other hand, diverges. However, since for some stimuli $I_2({\bf x})$ is
negative and for some others it is positive, the average of them all is finite,
as stated in Eq. (\ref{it}). The infinitely large discrepancy between Eqs.
(\ref{stli1}) and (\ref{stli2}) shows that for small $t$, the choice of any
one of these estimators is particularly meaningless.

As pointed out above, although this procedure is usually referred to
as a short time limit, the crucial step in deriving equations
(\ref{it} - \ref{stli2}) is to partition the set of all possible
responses into two subsets: one containing a single response, namely
the case $n = 1$, and the complementary one. The conditional probabilities
for the occurrence of the distinguished response ($n = 1$) is taken
proportional to a parameter $t$ which is supposed to be small.
Such a procedure with the response variable inspires the exploration of
an analogous partition in the set of locations.

\subsection{A spatial information density}

To find a well behaved measure of a location
specific information we now introduce the {\it local information}
$I^\ell({\bf x}_j)$, which quantifies how much can be learned
from the responses about whether the animal is or is not in ${\bf x}_j$.
In other words, we partition the set of possible locations into two subsets,
one containing position ${\bf x}_j$ and the complementary set
$\bar{{\bf x}}_j = \{{\bf x} / {\bf x}$ does not belong to region $j \}$.
Mathematically,
\begin{equation}
I^\ell({\bf x}_j) = \sum_{n = 0}^{+\infty} P(n)
\left\{ P({\bf x}_j | n) \log_2\left[ \frac{P({\bf x}_j | n)}{P({\bf x}_j)}
\right]  +  P(\bar{{\bf x}}_j | n) \log_2\left[ \frac{P(\bar{{\bf x}}_j | n)}
{P(\bar{{\bf x}}_j)} \right] \right\},
\label{binary}
\end{equation}
where $P(n)$ is the probability of the cell firing $n$ spikes no matter where,
$P({\bf x}_j)$ is the probability of visiting location ${\bf x}_j$,
$P(\bar{{\bf x}}_j) = 1 - P({\bf x}_j)$ is the probability of {\it not} being
in ${\bf x}_j$, $P({\bf x}_j| n)$ is the conditional probability of being
in ${\bf x}_j$ when the cell fired $n$ spikes,
$P(\bar{{\bf x}}_j | n)$ is the conditional probability of {\it not} being
in ${\bf x}_j$ while the cell fires $n$ spikes,
and can be obtained from
\begin{equation}
P({\bf x}_j | n) + P(\bar{{\bf x}}_j | n) = 1.
\end{equation}
Equation (\ref{binary}) defines a proper information, in the sense that
it is positive and additive.
It should be noticed that in contrast to the short time limit, in the
case of the local information, there is no preferred location to be
separated out. That is why we calculate as many $I^\ell({\bf x}_j)$ as
there are positions ${\bf x}_j$. As $j$ changes, however,
$I^\ell({\bf x}_j)$ refers to a different partition of the
environment. This means that one
should not average  the various $I^\ell({\bf x}_j)$.

In  parallel to the short time limit, we now make the area $\Delta$
of region ${\bf x}_j$ tend to zero. To do so, we assume that both
$P({\bf x}_j)$ and $P({\bf x}_j | n)$ arise from a continuous spatial
density $\rho$,
\begin{eqnarray}
P({\bf x}_j | n) &=&  \int_{{\rm region} \ j }
\rho({\bf x} | n) d{\bf x} \\
P({\bf x}_j) &=&
\int_{{\rm region} \ j } \rho({\bf x}) d{\bf x},
\end{eqnarray}
where
\begin{equation}
\rho({\bf x}) = \sum_{n = 0}^{+\infty} P(n) \rho({\bf x}_j | n).
\end{equation}
For $\Delta$ sufficiently small,
\begin{eqnarray}
P({\bf x}_j | n) &\approx& \Delta \ \rho({\bf x} | n) \nonumber \\
P({\bf x}_j) &\approx& \Delta \ \rho({\bf x}).
\label{quick}
\end{eqnarray}
The continuity of $\rho$ allows us to drop the sub-index $j$.
Expanding $I^\ell({\bf x})$ in powers of $\Delta$ it may be
seen that the first term is
\begin{equation}
I^\ell({\bf x}) = \Delta \sum_{n = 0}^{+\infty} P(n)
\left\{\rho({\bf x} | n) \log_2\left[\frac{\rho({\bf x} | n)}
{\rho({\bf x})} \right] + \frac{\rho({\bf x}) - \rho({\bf x}
| n)}{\ln 2} \right\} + {\cal O}(\Delta^2).
\label{idex}
\nonumber
\end{equation}
The local information is therefore proportional to $\Delta$, which means
that in the limit $\Delta \to 0$ it becomes a differential.
Equation (\ref{idex}) is completely analogous to (\ref{it}).
This behavior indicates that the density
\begin{equation}
i({\bf x}) = \frac{I^\ell({\bf x})}{\Delta}
\label{densidad}
\end{equation}
approaches a well defined limit when $\Delta \to 0$.
As pointed out earlier, $I^\ell({\bf x})$ is  conceptually different
from the full information $I$ defined in equation (\ref{inf}), and
for finite $\Delta$, there is no meaning in summing together
the $I_\ell({\bf x}_j)$ corresponding to different $j$. However, it
is easy to verify that when $\Delta \to 0$, the integral of
$i({\bf x})$ throughout the whole space coincides with
the full information $I$, when the latter is calculated in the limit
of very fine binning. Therefore, $i({\bf x})$ behaves as an
information spatial density. Moreover, in contrast to $I_1(x)$ and
$I_2(x)$, it derives from a properly defined information.


Equation (\ref{idex}) is the continuous version of
(\ref{binary}). It should be noticed, however, that in practice
one cannot calculate $\rho({\bf x} | n)$ from experimental data,
for finite time bins. If $\Delta$ is sufficiently small, and the
animal moves around with a given velocity, it just never remains
within ${\bf x}_j$ during the chosen time window. Nevertheless,
there is no inconvenient in giving a theoretical definition of
$\rho({\bf x} | n) = \lim_{\Delta \to 0} P({\bf x}_j | n) /
\Delta$, imagining one could perform the experiment placing the
animal in ${\bf x}_j$ and confining it there throughout the whole
time interval. In order to bridge the gap between the theoretical
definition of $\rho({\bf x} | n)$ and the actual possibility of
measuring an information spatial density with freely moving
animals, we now take equation (\ref{idex}) and calculate its
short time limit. The result is
\begin{equation}
I^\ell({\bf x}) = t \Delta \rho(x) \left\{r(x) \log_2
\left[ \frac{r(x)}{\langle r \rangle} \right] + \frac{\langle r \rangle -
r(x)}{\ln 2} \right\} + {\cal O}(\Delta^2) + {\cal O}(t^2).
\label{linear}
\end{equation}
This same expression is obtain if one starts with the full definition
(\ref{binary}) and first calculates the limit of $t \to 0$ and subsequently
makes $\Delta \to 0$.
Comparing equation (\ref{linear}) with (\ref{stli1}) it is clear that
\begin{equation}
I^\ell({\bf x}) = t P(x) I_1({\bf x}) + {\cal O}(\Delta^2) + {\cal O}(t^2).
\end{equation}
We therefore conclude that in the short time limit, the position
specific surprise coincides with the local information (multiplied
by the probability of occupancy). This gives the position specific
surprise $I_1$ a different status than $I_2$ or any combination of
the two: even though in a general situation $I_1$
is not additive, when the number of stimuli is very large (or the
binning very fine, if the stimuli are continuous) it coincides with
a well defined quantity, namely, the local information.
It should be kept in mind, however,
that such a correspondence between $I^\ell({\bf x})$ and $I_1(\bf x)$
is only valid in the short time limit---or, more precisely, when
computing the information provided by one spike.


\section{Data analysis}

In the present section we evaluate $I^\ell({\bf x}_j)$ as a function of
${\bf x}_j$ using electrophysiological data recorded from rodents performing
a spatial task. In Subsection 3.1 the experimental procedure is explained,
and later, in 3.2 we show that different brain regions use different
coding strategies in the representation of space.


\subsection{Experiment design}

Nine young adult Long-Evans rats were tested during a
forced-choice and a spatial working memory task. Both tasks were
performed in an 8 arm radial maze, each arm containing a small
amount of chocolate milk in its distal part. In addition, the
proximal part of each arm could be selectively lowered, thereby
forbidding the entrance to that particular arm (for details see
Leutgeb {\it et al.}, 2000). In the forced choice task, the animal
was placed in the center of the maze, while the entrance to a
single arm was available. The other seven arms were kept at a
lower level. After the animal had entered into the available arm
and taken its food reward, a second arm was raised, and the
previous one was lowered. The procedure was continued by allowing
the animal to enter just one arm at a time with no repetitions.
The session ended when the rat had taken the milk of all eight
arms. A pseudo-random sequence of arms was chosen for each trial.
The beginning of the  working memory task was identical to the
forced choice one, until the animal had entered the fourth arm in
the sequence. At this point, all arms of the maze were raised, and
the rat could move freely. However, only the four new arms still
contained food reward. The session continued until the animal had
taken all the available chocolate milk or for a maximum of 16
choices. Re-entries into previously visited arms of the maze were
counted as working memory errors, since in principle, the animal
should have kept in mind that in that arm, the food had already
been eaten.

Septal and hippocampal cells were simultaneously recorded during
both of the tasks (for recording details see Leutgeb 2000). A
head-stage held the FET amplifiers for the recording electrodes
as well as a diode system for keeping track of animals position.
Single units were separated using an on-line and off-line
separation software. Units are then identified according to their
anatomical location and the characteristics of the spikes.
Hippocampal pyramidal cells and interneurons as well as lateral
and medial septal cells were identified. Animals were tested
either in standard illumination condition or in darkness.

\subsection{Results and discussion}

In order to compute $I^\ell({\bf x}_j)$ from the experimental
data, both $r({\bf x}_j)$ and $P({\bf x}_j)$ are needed for each
position. In order to compute $r({\bf x})$ the total number of
spikes fired in location ${\bf x}_j$ is divided by the total time
spent there. The a priori probability $P({\bf x}_j)$ of visiting
the spatial bin $j$ was obtained by the ratio of the time spent
in ${\bf x}_j$ and the total duration of the trial. The
computation of mutual information typically introduces an upward
bias due to limited sampling. Therefore, a correction has been
applied, in order to reduce this overestimation, as suggested in
Panzeri and Treves (1996). In our case,  the first order
correction for equations (\ref{it}) and (\ref{binary}) can be
derived analytically
\begin{equation}
I_{\rm corr} = I -  \frac{t (N - 1)}{ 2 T \ln 2},
\label{limsam}
\end{equation}
where $N$ is the number of positions ${\bf x}_j$ in which the
environment has been binned and $T$ is the total duration of the
recording. Throughout the paper, when specifying experimental
data, we always give the corrected value of $I$---although, for
simplicity of notation, we drop the sub-index ``corr''.
\begin{table}[h]
\begin{center}
\begin{tabular}{||l|r|r|r|r|r||} \hline
neuron &  number  & $\langle \langle r \rangle \rangle$ &
$\sigma_{\langle \langle r \rangle \rangle}$ & $\langle I \rangle
/ t$
& $\sigma_{\langle I \rangle} / t$ \\
& of units &  spikes / sec. & spikes / sec. & bits / sec. &  bits
/ sec.
\\ \hline
HP & 114 & 0.99665 & 1.2353 & 0.43851 &0.39535 \\ \hline HI & 21
& 15.228 & 6.9618 & 0.81994 & 0.85995 \\ \hline LS & 327 & 5.0732
& 7.7119 & 0.25274 & 0.33575  \\ \hline MS & 34 &  10.971 &
12.177 & 0.27802 & 0.46498 \\ \hline
\end{tabular}
\end{center}
\caption{\normalsize Statistic corresponding to the whole
poulation of recorded neurons, where $\langle \langle r \rangle
\rangle$ is the mean firing rate averaged throughout the maze,
and over all cells, $\sigma_{\langle \langle r \rangle \rangle}$
is the average quadratic deviation from the mean, $\langle I
\rangle / t$ is the mean information rate, averaged over cells,
and
 $\sigma_{\langle I \rangle}/t$ its mean quadratic deviation. HP stands
for pyramidal cells in the hippocampus, HI for interneurons in
the hippocampus, while LS  for units in the lateral, and MS for
unuits in the medial septum.} \label{t1a}
\end{table}
Table \ref{t1a} summarizes the overall statistics of our
experimental data. The values of $I/t$ have been calculated in
the short time limit (\ref{it}) and further subtracting the
correction (\ref{limsam}).

\begin{figure}[htbf]
\begin{center}
\leavevmode \epsfxsize=8.0truecm \epsffile{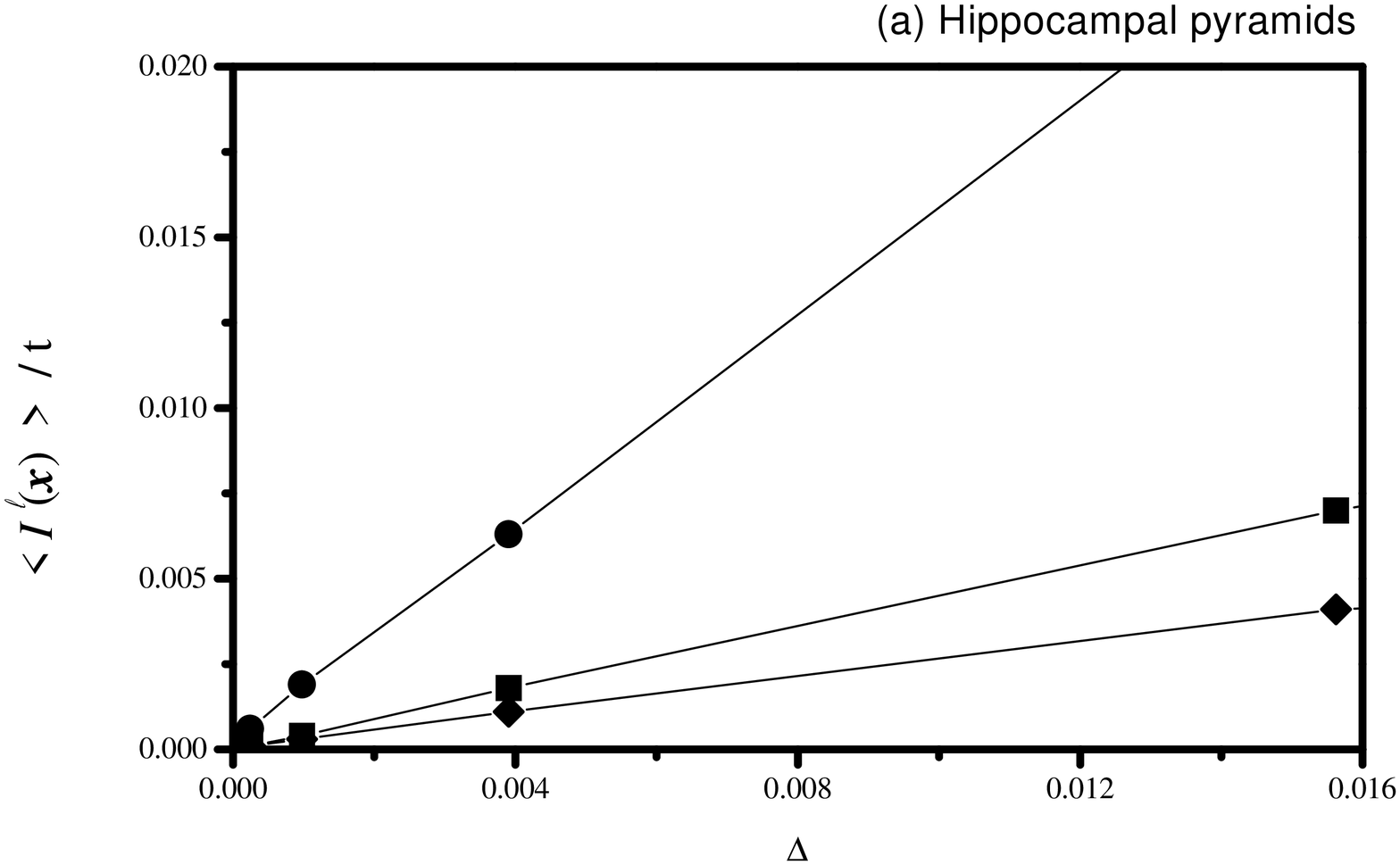}
\epsfxsize=8.00truecm \epsffile{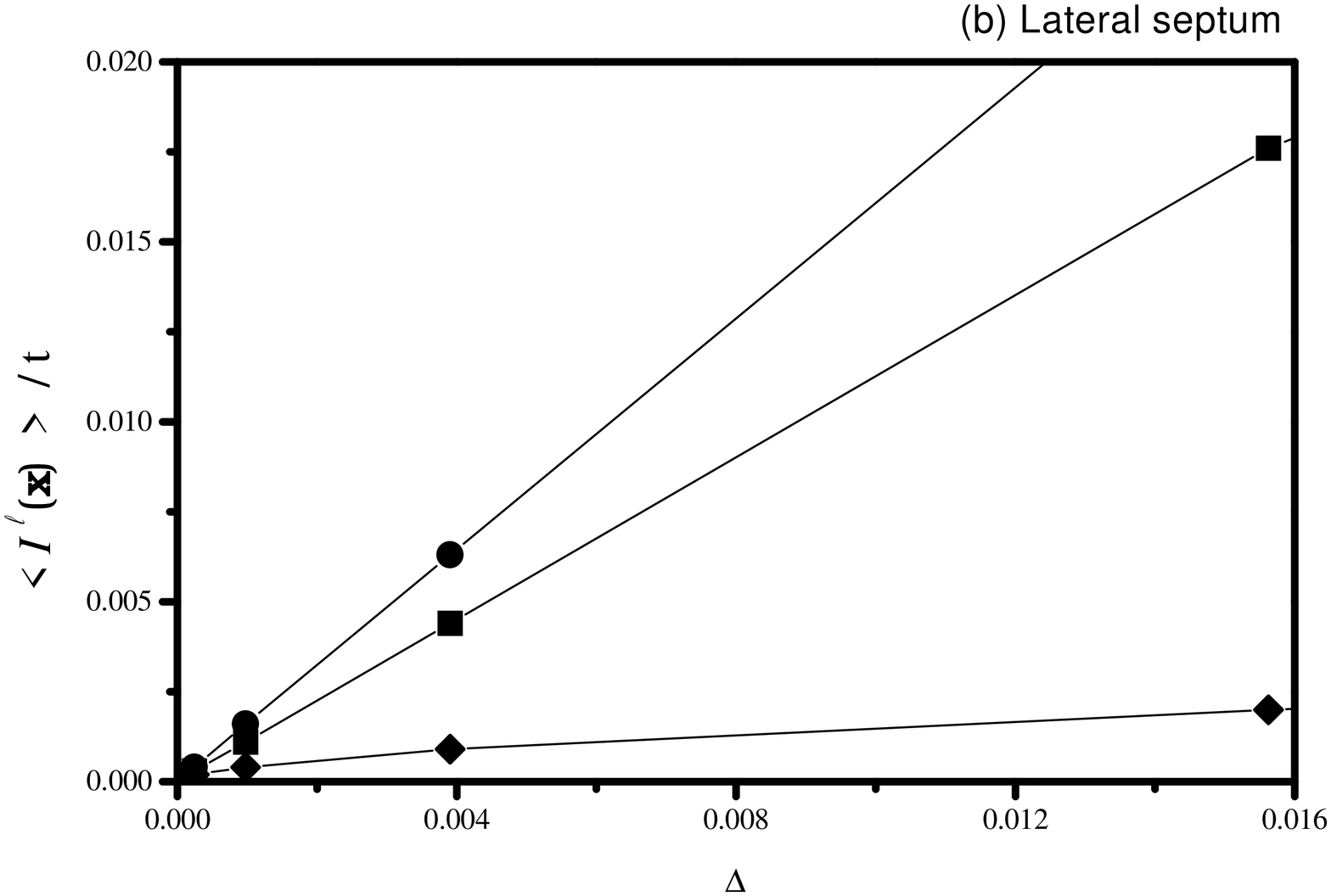}
\end{center}
\caption{Mean local information rate---defined in equation
(\ref{media})---as a function of the area $\Delta$ of each bin.
Three pyramidal cells in the hippocampus are shown in (a) and
 three  cells in lateral septum cells appear in (b). All cells carry an
appreciable amount of information when compared to other cells of
the same type. Both (a) and (b) contain one unit with a high
firing rate, an intermediate one and a low frequency cell.}
\label{f1}
\end{figure}

The proportionality between $I^\ell({\bf x})$ and $\Delta$
(see Eq.~\ref{linear}) was based
on the assumption that the conditional probabilities $P({\bf x}_j | n)$
emerged from a continuous density $\rho({\bf x} | n)$. In order to
verify whether such a supposition actually holds, we evaluated
$I^\ell({\bf x}_j)$ from our experimental data,
for different values of the area $\Delta$.
In Figure \ref{f1} we show
a spatial average of our results, namely
\begin{equation}
\langle I^\ell({\bf x})\rangle = \frac{1}{N} \sum _j I^\ell({\bf x}_j),
\label{media}
\end{equation}
where $N$ is the total number of positions $j$ in which the maze
has been binned. Clearly, $N = $ total area $ / \Delta$. The local
information $I^\ell({\bf x}_j)$ has been evaluated in the short
time limit. It is clearly seen that in all cases, the local
information grows linearly with $\Delta$, as predicted by
Eq.~(\ref{idex}).

\begin{figure}[htbf]
\begin{center}
\leavevmode \epsfxsize=10.0truecm \epsffile{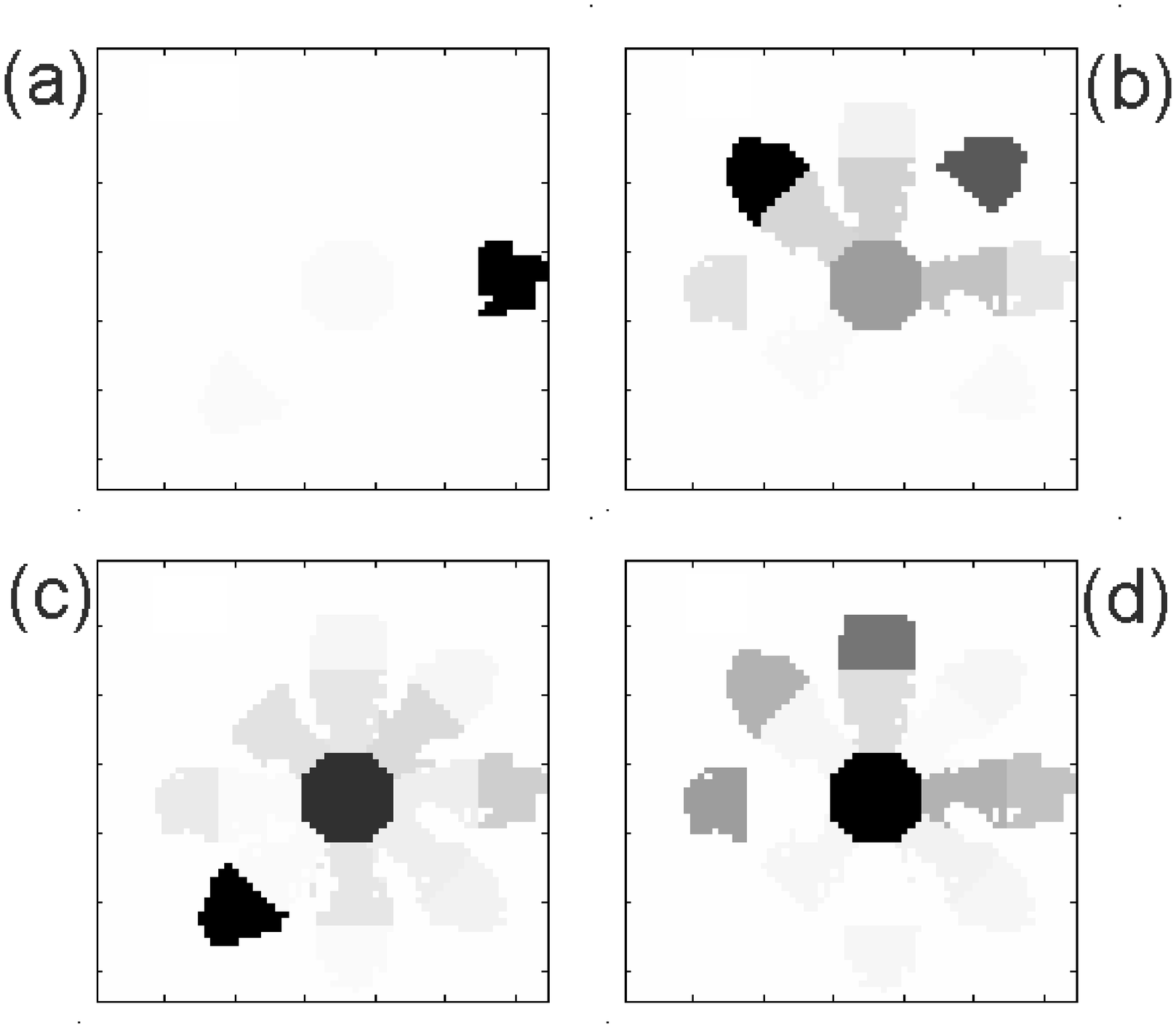}
\end{center}
\caption{Local information distributions corresponding to the
firing densities of Figure \ref{f0}. In each case, the density of
the color is proportional to the number of bits per unit time, as
a function of space. } \label{f2}
\end{figure}

We therefore build
the local information maps, for all the cells recorded. In other
words, we calculate $i({\bf x}_j)$ for all the positions ${\bf x}_j$.
We have restrained ourselves from  going into a too fine binning, however, in order to
avoid limited sampling problems. In Figure \ref{f2} we show the information
maps corresponding to the firing distributions of Figure \ref{f0}. The
hippocampal pyramidal cell in (a) is only informative at the same
location where the cell fires. In this particular case, the  intuition
seems to be justified: the cell codes for a single position, and does
so by increasing its firing rate at that location. However, the other three
cases show that the neuron may well provide information not only where
it fires most, but also where it fires least. In particular, the
hippocampal interneuron in (b), which tends to fire all over the maze,
is particularly informative in the upper-left and upper-right end points,
where it remains almost silent. In cases (c) and (d) the cells
provide information both where there is a high and a low firing rate.
As a consequence, we conclude that if a cell has a distributed coding
scheme (as opposed to a very local one, more typical of hippocampal place
cells), the information map may well not coincide with the
firing rate one. In this sense, one should beware not to judge cells
with a distributed firing pattern as non-informative. If such a pattern
is stable and reproducible, covering a wide range of firing rates, the
neuron may well be providing spatial information.

Could a quantitative analysis of the coding strategies
of hippocampal pyramidal cells, and neurons in the lateral
septum be given? We have not considered hippocampal interneurons nor
medial septum cells since in these two cases, we do not have enough
statistics to draw conclusions. In addition,
on average, they are less informative (see table \ref{t1a}).

\begin{figure}[htbf]
\begin{center}
\leavevmode \epsfxsize=8.0truecm \epsffile{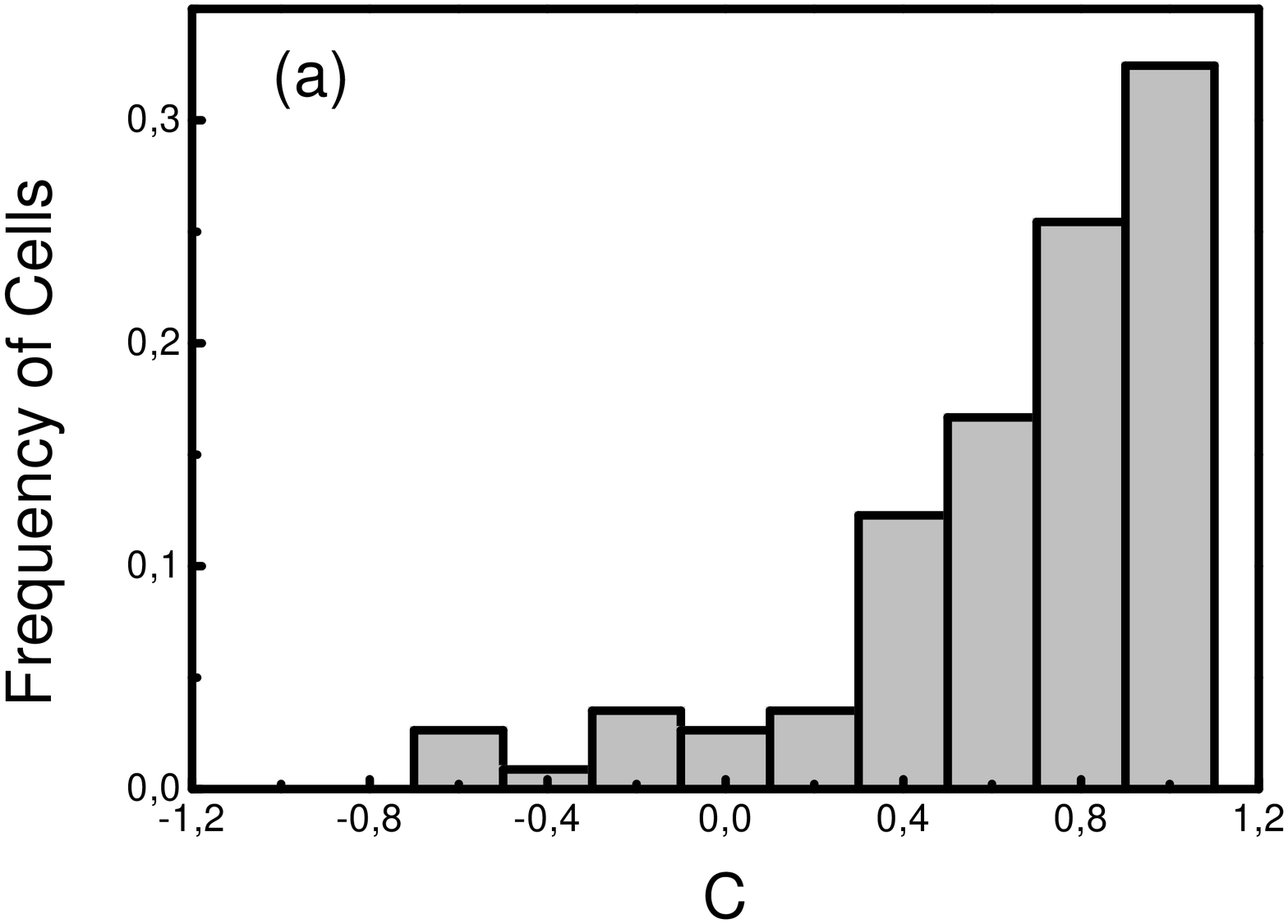}
\epsfxsize=8.0truecm \epsffile{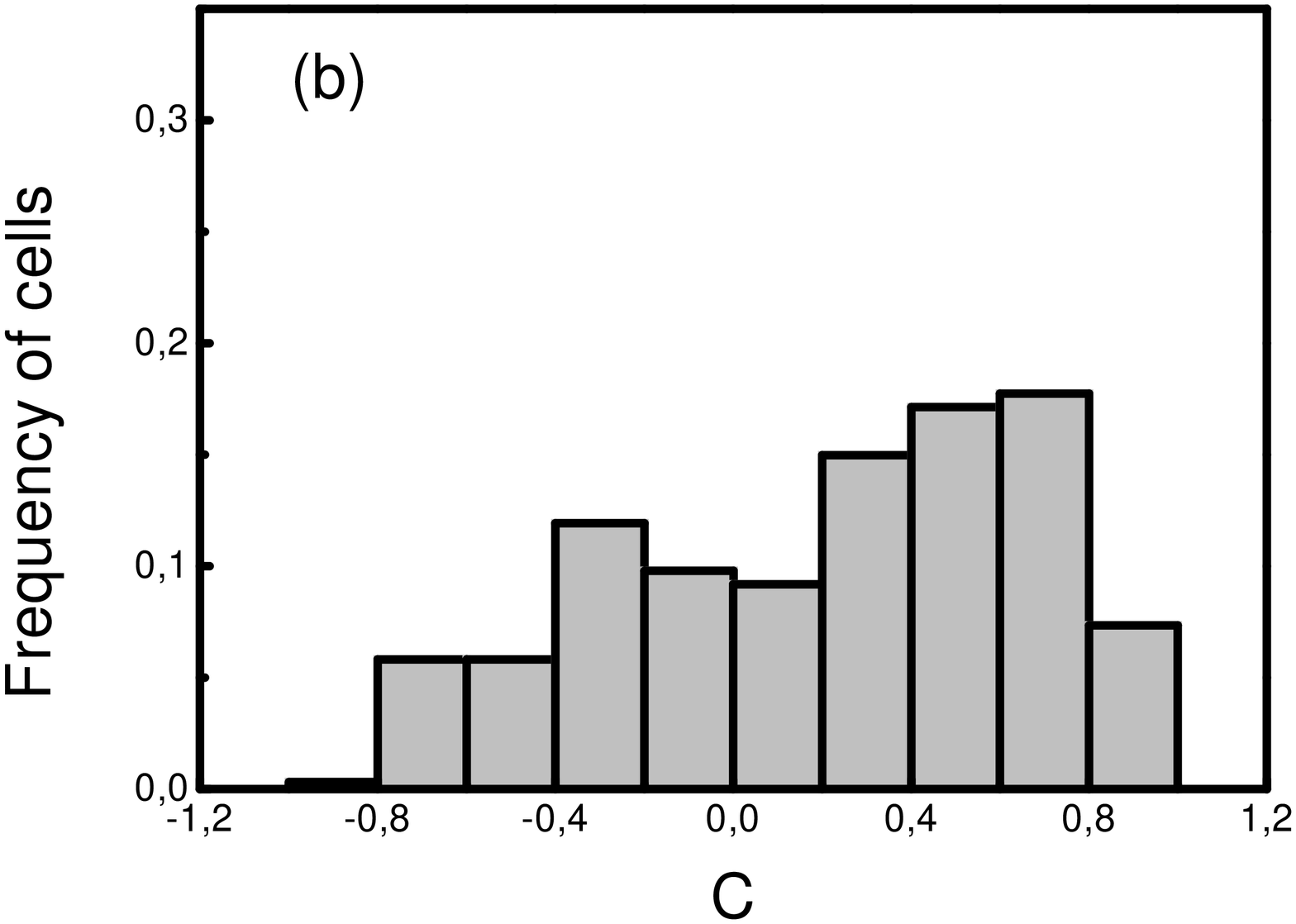}
\end{center}
\caption{Frequency distribution of the correlation $C$ between
the information and firing rate spatial distributions, for (a)
114 pyramidal cells in the hippocampus, and (b) 327 neurons in
the lateral septum.} \label{f3}
\end{figure}

One possible measure of the degree of localization of the coding scheme
is given by the correlation between the information maps and the firing
rates distributions (that is, between the graphs of figures \ref{f0} and
\ref{f2}). We evaluate  the Pearson correlation coefficient between
 the two maps, i.e.
\begin{equation}
C = \frac{ \sum _j \left[ I^\ell({\bf x}_j) - \langle
I^\ell({\bf x}_j) \rangle _j \right] \ \left[ r({\bf x}_j) -
\langle r \rangle \right]}{ N \left [ \sum _j I^\ell({\bf x}_j)\right]
\left[ \sum _j r({\bf x}_j)\right]},
\end{equation}
where $\langle I^\ell({\bf x}_j)\rangle_j$ is the spatial average of
the local information.
Thus, $C$ is equal to 1 if $I^\ell({\bf x}_j) - \langle I^\ell({\bf x}_j)
\rangle_j$ is proportional to $r({\bf x}_j) - \langle r \rangle$, and
takes the value -1, if the proportionality factor is negative.
Notice that there is one such $C$ for every single cell.
In Figure \ref{f3} we show the frequency distribution of the correlation
$C$ for (a) the $114$ pyramidal cells recorded in the hippocampus and (b)
the $297$ units recorded in the lateral septum. It may be seen that
pyramidal cells tend to have larger values of the correlation $C$,
indicating that they tend to provide information in the same locations
where they fire most. In other words, the code in the hippocampus
can be characterized as localized, as is well known, giving rise to place
cells and place fields. In contrast, septal cells have a somewhat
more symmetrical distribution around zero. If $C \approx 0$, then the cell
gives as much information in those locations where it fires most, as
where it remains silent (or, more precisely, where it fires less than
its average spontaneous rate). A negative value of $C$ indicates that
the cell is most informative in the locations  where it does not
fire (Figures \ref{f0} (b) and \ref{f2} (b) provide an example
from an hippocampal interneuron). As stated in Table \ref{t1a}, hippocampal
pyramidal cells are more informative than lateral septal cells. The point we want to
stress is that the lateral septum follows
a different coding strategy: cells do not specialize in a particular region of space,
but rather, respond with a complex, distributed firing pattern all over the place.

\section{Conclusions}

The aim of the present work was to present a characterization of the way the
information provided by neural activity distributes among the elements
of a given set of stimuli. In our case, the stimuli were the different
positions an animal can be located, within its environment.
We defined the local information
$I^\ell(s)$, namely the information provided by the responses of
whether the stimulus  is or is not $s$. In other words, it is the mutual
information between the responses and a reduced set of stimuli,
consisting of only two elements: stimulus $s$, and its complement.
In contrast to other quantities introduced previously, this is a well
defined mutual information which can be employed in the short time limit.
In fact, other possible definitions have some drawbacks;
for example, the position specific surprise has the disadvantage of not
being additive. From the theoretical point of view it is therefore not
a very sound candidate for quantifying the information to be associated
to each stimulus. The position specific information, on the other hand,  may
not be positive and diverges for $t \to 0$, thus making its application to
actual data quite cumbersome.

In this paper we have studied the properties
of $I^\ell$ in the particular situation were the stimuli arise from a
continuous variable (as position in space) which has an underlying metric.
In this case, the binning that transforms the continuous variable
(in our case ${\bf x}$) into a discrete set $({\bf x}_j)$ may be chosen,
in principle, at will. When working with real data, however, the size of
the bins is determined by the amount of data, since the mean error in
the calculation of the mutual information is linear in the number of
bins (see equation (\ref{limsam})).

We have shown analytically that when the size of the bin goes to zero,
the local information is proportional to the bin size. This means that
$I^\ell({\bf x})/\Delta$ behaves as an information spatial density, in that
it tends to a constant value when $\Delta \to 0$, and its integral all
over space coincides with the full information $I$. We point out that these
two properties hold only in the limit of $\Delta \to 0$, whereas
the position specific surprise and the position specific information
fulfill equation (\ref{parc}) for any size of the bins.

We have also shown that in the short time limit and for $\Delta \to 0$,
the local information coincides with the position specific surprise,
multiplied by the probability of occupancy. This result may seem puzzling,
since $I_1$ is known not to be additive, while additivity is guaranteed
in $I^\ell$. However, it should be noticed that the additivity of
$I^\ell$ is more restricted than the one desired for $I_1$. If the
position specific surprise were to be additive, $I_1$ would obey the
relation
\begin{equation}
I_1({\bf x}_{j_1}, {\bf x}_{j_2}) = I_1({\bf x}_{j_1}) + I_1({\bf x}_{j_2}
| {\bf x}_{j_1}),
\label{lio1}
\end{equation}
where $I_1({\bf x}_{j_1}, {\bf x}_{j_2})$ is the information
provided by the responses about two particular results of the
measurement of the stimulus. As shown by De Weese and Meister
(1999), $I_1$ does not follow equation (\ref{lio1}). The local
information, on the other hand, does fulfill the condition
\begin{equation}
I^\ell({\bf x}_a, {\bf x}_b) = I_1({\bf x}_a) + I_1({\bf x}_b
| {\bf x}_a),
\label{lio2}
\end{equation}
where ${\bf x}_a$ and ${\bf x}_b$ may only be ${\bf x}_j$ or
$\bar{{\bf x}}_j$, and $I^\ell({\bf x}_a, {\bf x}_b)$ represents a true
mutual information, between the set of responses and the two sets
$\{{\bf x}_j,\bar{{\bf x}}_j\}$ (one set for each measurement).
Additivity for $I_1$ requires additivity for any choice
of ${\bf x}_{j_1}$ and ${\bf x}_{j_2}$ in (\ref{lio1}), while the
possible values of ${\bf x}_a$ and ${\bf x}_b$ are much more restricted
in (\ref{lio2}). One should therefore not mistrust the fact that
$I_1$ does not obey a very demanding additivity condition, while
$I^\ell$ fulfills a quite relaxed one.

The local information, as defined here, allows the
characterization of the spatial properties of the information
conveyed by a given cell. Just by measuring the mutual
information between a given neuronal response and the set of
possible locations, one sees that there are cells (both in the
hippocampus and in the lateral septum) that provide an appreciable
amount of information without actually having a place field. By
means of the local information, it is possible to draw a spatial
information density which may, in these non-typical cases, differ
significantly from the rate distribution. In the last section we
have shown that the local information can be easily calculated
from experimental data, and that it can actually be used to
characterize the coding strategy of different cell types. In
particular, we saw that hippocampal place cells tend to provide
spatial information in the same places where they fire, whereas
lateral septal neurons use a more distributed coding scheme. This
result is in agreement with the different goals in the encoding of
movement related quantities in both regions, as described
recently in Bezzi {\it et al.} (2000).


\section*{Acknowledgements}
We would like to thank Bill Bialek and Alessandro
Treves for very useful discussions.
This work was supported by the Human Frontier Science Program,
Grant No. RG 01101998B.

\newpage


\section*{References}

\begin{itemize}

\item[-] Bezzi, M., Leutgeb, S., Treves, A., \& Mizumori, S. J. Y.
(2000). Information analysis of location-selective cells in
hippocampus and lateral septum. {\it Society for Neuroscience
Abstracts}, 26, 173.11.

\item[-] De Weese, M. R., \& Meister, M. (1999). How to measure the
information gained from one symbol. {\it Network}, {\bf 10},
325-340.

\item[-] Jakab, R., \& Leranth, C. (1995). In {\it The Rat Nervous
System}, 2nd Ed., 405-442; edited by G. Paxinos. New York:
Academic Press.

\item[-] Leutbeb, S., \& Mizumori, S. J. (2000). Temporal
correlations between hippocampus and septum are controlled by
enviromental cues: evidence from parallel recordings. Submitted.

\item[-] O'Keefe, J., \& Dostrovsky, J. (1971). The hippocampus as a
spatial map: Preliminary evidence from unit activity in the
freely moving rat. {\it Brain Res.}, {\bf 34}, 171-175.

\item[-] Panzeri, S., Biella, G., Rolls, E. T., Skaggs, W. E., \&
 Treves, A. (1996). Speed, noise, information and the graded nature of
neuronal responses. {\it Network}, {\bf 7}, 365-370.

\item[-] Panzeri, S. \& Treves, A. (1996). Analytical estimates
of limited sampling biases in different information measures. {\it
Network}, {\bf 7}, 87-107.

\item[-] Skaggs, W. E., McNaughton, B. L., Gothard, K., \& Markus, E.
(1993). An information theoretic approach to deciphering the
hippocampal code, pp 1030 - 1037 in {\it Advances in Neural
Information Processing Systems} Vol 4, eds. S. J.  Hanson, J. D.
Cowan \& Giles, C. L. Morgan. Daufmann: San Mateo, CA.

\item[-] Swanson, L. W., Sawchenko, P. E. \& Cowan, W. M. (1981).
Evidence for collateral projections by neurons in Ammon's horn,
the dentate gyrus, and the subiculum: a multiple retrograde
labeling study in the rat. {\it J. Neurosci.},{\bf 1}, 548-559.

\item[-] Tsodyks, M. (1999). Attractor Neural Network Models of
Spatial Maps in Hippocampus. {\it Hippocampus}, {\bf 9}, 481-489.

\item[-] Zhou, T., Tamura, R., Kuriwaki, J., \& Ono, T., (1999).
Comparison of medial and lateral septum neuron activity during
performance of spatial tasks in rat. {\it Hippocampus}, {\bf 9},
220-234.

\end{itemize}



\end{document}